\documentstyle[12pt,epsf,draft]{article}

\def\sign{\mathop{\rm sign}\nolimits}
\makeatletter
\def\eqalign#1{\null\vcenter{\def\\{\cr}\openup\jot\m@th
  \ialign{\strut$\displaystyle{##}$\hfil&$\displaystyle{{}##}$\hfil
      \crcr#1\crcr}}\,}
\makeatother

\begin{document}

\author{V.~M.~Kontorovich\\
{\em Institute of Radio Astronomy}\\
{\em Krasnoznamennaya 4, 310002, Kharkov, Ukraine}\\
{\em E-mail: vkont@ira.kharkov.ua}\\
and\\
S.~F.~Pimenov\\
{\em Department of Mathematics and Statistics}\\
{\em Monash University, Clayton, Victoria 3168, Australia}\\
{\em E-mail: sfp@brain.physics.swin.oz.au}
{\em (new affiliation)}}
\title{Exact Solution of the Kompaneets Equation for Strong Explosion in
Medium with Inverse-Square Decreasing Density}
\date{February 3, 1998}
\maketitle

\begin{abstract}
In the framework of the Kompaneets approximation the propagation
of a shock front (SF) in inhomogeneous medium with the power-law
density decrease at the exponent $n=2$ is investigated. For this
important case corresponding to outer regions of the solar and
stellar coronas an unexpectedly simple exact solution clearing a
structure of the general solution for an arbitrary monotonic
density of medium is found. Our results for plane-layered medium
are compared to the Korycansky one for off-centre explosion in
radially stratified medium.  The relation between the solutions in
these media are found and a new exact solution for a noncentral
explosion in the case of density singularity on the finite radius
is received.
\end{abstract}

\subsection*{Introduction}

As really seen from the well-known Sedov--Tailor formula in the
homogeneous medium \cite{ll,s}
\begin{equation}
\label{1}
R=[E_{0}t^2/\rho _{0}]^{1/5}
\end{equation}
for the dependence of the radius $R$ of a shock front on time $t$
from the moment of explosion, the explosion energy $E_0$ and the
medium density $\rho _0$, the SF would move faster towards the
direction of density decrease if explosion occurs in inhomogeneous
medium. However, for inhomogeneous medium the problem becomes more
complicated, being not self-similar, and requires significantly
more complicated calculations even for computational analysis (see
the review and papers \cite{bs,kl,hn} and the bibliography
therein).

Assuming the uniform pressure along the envelope behind the SF
A.~S.~Kompaneets proposed the equation \cite{kom} (see
Eq.\ (\ref{3}) below) for strong explosion in an exponential
atmosphere. In that case he was showed that the SF was accelerated
and could get infinity for a finite time. (Surely, the
approximation fails earlier and account for a non zero density
at infinity \cite{sf} restricts values of the SF velocity too,
i.e., there is no controversy with physics or common sense.) For
densities decreasing not so fast, namely, for the power-law
density dependence $\rho (z) \propto z^{-n}$, often met in
astrophysical objects, the behaviour of the SF is defined by the
exponent $n$ \cite{s,kor,kp6}. The asymptotically exact solution
for the head part of the SF moving towards the direction of
density decrease indicates that the Sedov--Tailor SF deceleration
is down when the density gradient (or $n$) is up\footnote{Note
that at $n = 3$ the SF moves at a constant velocity. For $n$
greater than 3 the SF is accelerated and since $n = 5$ it gets
infinity for a finite time \cite{kor,kp6,kp}. The solution
qualitatively becomes similar to the Kompaneets one in an
exponential atmosphere.}.  Particularly, it is confirmed by the
exact solution of the Kompaneets equation discussed below for
$n=2$ \cite{kp5} corresponding to the wind regions in outer parts
of the solar and stellar coronas (however, with the wind motion
not taken into account). This solution having an extremely simple
form is analysed in this publication.

Following Kompaneets \cite{kom}, instead of the real time $t$
we define the new temporal variable $y$ as
\begin{equation}
\label{2}
dy=dt[E_{0}\lambda (\gamma ^{2}-1)/(2\rho _{0}V(t))]^{1/2},
\end{equation}
including into it the inverse square root of the time dependent
volume $V(t)=\pi \int r^{2}\,dz$. After substitution the
coefficients of the equation for the SF $r=r(y,z)$ do not depend
on time any more but only on the coordinate $z$. Along this
coordinate the density changes as $\rho(z)=\rho_0/(z/z_0)^2$.
Here $\rho_0$ is the unperturbed density in the point of explosion
$z_0$. Some another relations important for applications but not
for the current formal consideration are presented in
Appendix 1. The Kompaneets equation for the shock front can be
written as:
\begin{equation}
\label{3}
(\partial r/\partial y)^{2}\varphi (z) -
[(\partial r/\partial z)^{2}+1] = 0.
\end{equation}
Here $\varphi (z)=\rho (z)/\rho _{0}$.  In the case of $n=2$ the
solution could be immediately written, however, we will consider
it as a particular case of the general situation to clarify its
structure in more general cases of astrophysical interest (change
of the exponent $n$ from 16 to 6 and 2 in the solar corona; $n=3$
in active nuclei of galaxies).

The plan of this paper is the following. In the beginning we give
the quadrature parametric solution of the Kompaneets equation
(\ref{3}) (see \cite{kom}) for an arbitrary density distribution
of the medium using a supplementary function $\xi (y,z)$ and an
arbitrary function $\mu (\xi)$.  Regarding the medium at the
initial stage of expansion as homogeneous we conclude from
comparison to the Sedov solution that $\mu$ is to be zero (see
\cite{kom}). At $\mu=0$ for the inverse-square law of the density
decrease we find an explicit solution in two regions adjoined to
the opposite leading points of the SF. In this two regions the
solution consists of two non-adjoined segments of the same sphere.
In the intermediate region the function $\mu$ is not zero as
well as in the Silich \& Fomin case \cite{sf}, where the density
decreases exponentially to some constant value. The solution obtained
in the intermediate region complements the SF to a full sphere. It
is shown that at short times ($y \to 0$) the intermediate region
disappears. The transformation is also found of this solution in a
plane-layered medium to that for an off-centre explosion in the
radially stratified medium with the density distribution having a
singularity at a sphere surface with a finite radius, i.e., one
more new solution is obtained. (Preliminary comparison is carried
out to the Korycansky solution \cite{kor} for a power-law
radially-stratified medium). An extremely simple form of the solution
obtained admits explicit calculations which can be important for
general analysis of propagation of strongly non-linear waves in an
inhomogeneous media. A square-root singularity at the leading
points is revealed as an inherent feature of $\xi$ for an
arbitrary density dependence on coordinates.  This feature allows
some conclusions to be made about SF properties. The fact is that
different representations of the solution is required in different
regions. Probably, in the vicinity of every leading point we deal
with the only non-linear wave whereas contribution of both of
these waves (non-linear ``interference'') should be matched by
choosing the function $\mu$ in the intermediate region.

\subsection*{Initial condition}

The complete integral of Eq.\ (\ref{3}) dependent on two arbitrary
constants (according to the number of independent variables) can be
found via separation of variables (see, for example, \cite{kk}).
Regarding the integration constant $\mu $ as an arbitrary function
of the separation constant $\xi $ and substituting the latter with
the function of two independent variables $\xi (y,z)$, which can
be derived from the condition \cite{kk}
\begin{equation}
\label{5}
dr/d\xi =0,
\end{equation}
we find the following expression for the general integral:
\begin{equation}
\label{0.1}
r=-\sign(z-z_{0}) \int_{z_0}^zdx\sqrt{\xi ^2\varphi
(x)-1}+\xi y+\mu (\xi ).
\end{equation}
This general integral and the condition (\ref{3})
for the function $\xi $ give its implicit expression via $z$ and $y$
\begin{equation}
\label{0.3}
y=\sign(z-z_{0}) \int_{z_0}^z
dx\frac{\xi \varphi (x)}{\sqrt{\xi ^2\varphi (x)-1}}
-\mu^{\prime },\quad \frac{d\mu }{d\xi }\equiv \mu ^{\prime },
\end{equation}
from where for the solution describing the SF we have
\begin{equation}
\label{0.4}
r=\sign(z-z_{0}) \int_{z_0}^z\frac{dx}{\sqrt{\xi ^2\varphi
(x)-1}}+\mu -\xi \mu ^{\prime }.
\end{equation}
At $y\to 0$, $z\to z_0,$ $r\to 0$,
\begin{equation}
\label{0.5}
r\approx \frac 1{\sqrt{\xi ^2-1}}\,\vert z-z_0 \vert
+\mu -\xi \mu ^{\prime },\quad
y\approx \frac \xi {\sqrt{\xi ^2-1}}\, \vert z-z_0 \vert -\mu ^{\prime },
\end{equation}
or
\begin{equation}
\label{0.7}
\left( \frac{r-\mu +\xi \mu ^{\prime }}{{z-z_0}}\right) ^2\approx \frac
1{\xi ^2-1},\quad
\left( \frac{y+\mu ^{\prime }}{z-z_0}\right) ^2\approx  \frac{\xi ^2}{\xi
^2-1}.
\end{equation}
Generally, the supplementary functions $\mu$ and $\xi$ are considered not
expanded. Excluding $\xi ^2$ gives:
\begin{equation}
\label{0.9}
\left( \frac{r-\mu +\xi \mu ^{\prime }}{z-z_0}\right) ^2+1=\left( \frac{
y+\mu ^{\prime }}{z-z_0}\right) ^2.
\end{equation}
It is obvious that approaching the Sedov limit
\begin{equation}
r^{2}+ (z-z_{0})^{2} = y^{2}
\end{equation}
is possible only if (disregarding possible though non-essential
shift of initial moment of explosion)
\begin{equation}
\label{0.10}
\frac{\xi \mu ^{\prime }-\mu }r\to 0,\quad r\to 0;\quad \frac{\mu
^{\prime }}y\to 0,\quad y\to 0.
\end{equation}
Finally, for this limiting case of medium homogeneous at small distances the
following relations occur
\[
\mu ^{\prime }=0,\quad \mu =0.
\]
We will see below that the situation is more complicated, however,
for while we can use the relations derived for the exponent of
interest $n=2$.

\subsection*{Solution for $n = 2$ in the vicinity of leading points}

\begin{figure}
\centerline{\epsfbox{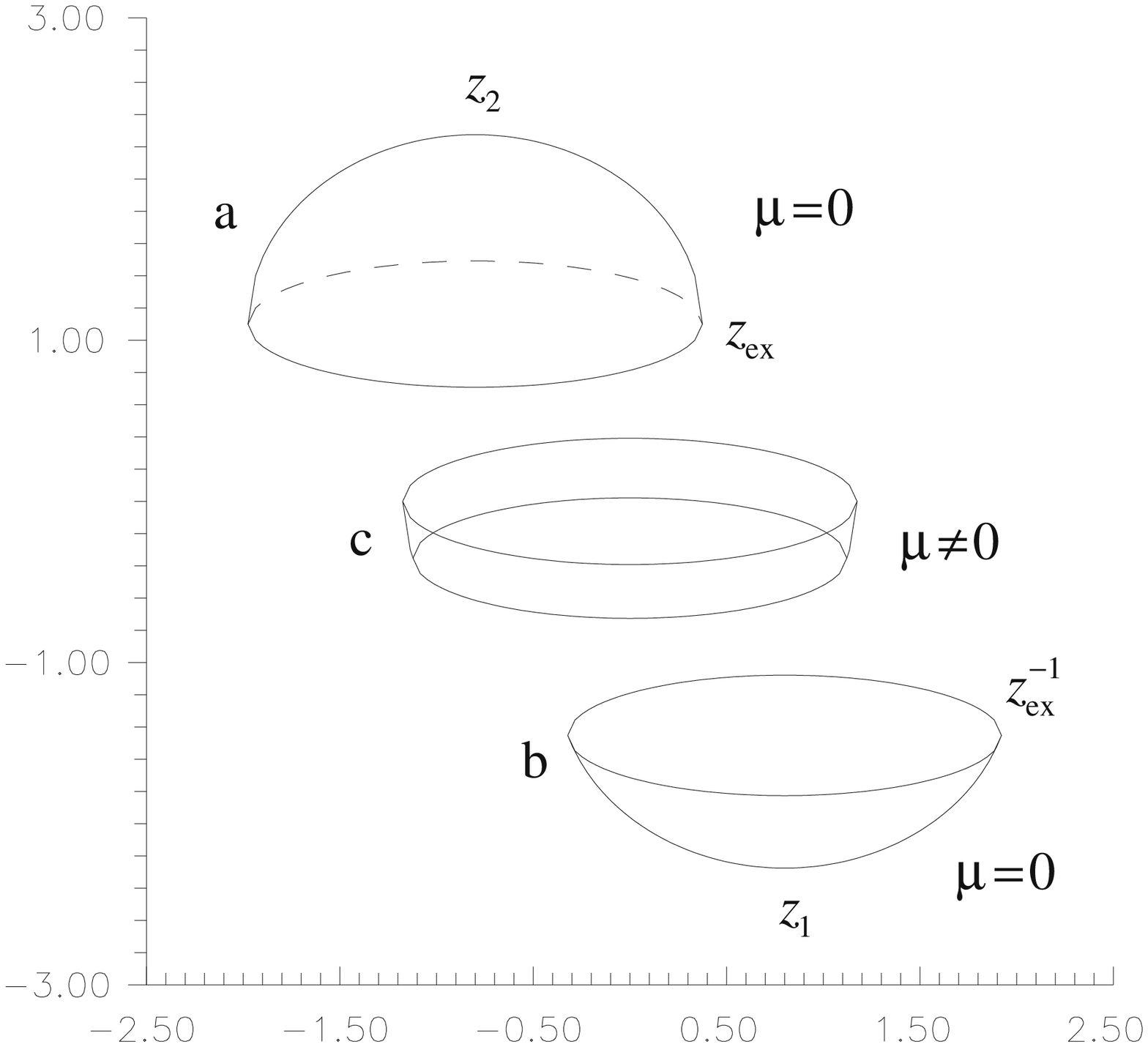}}
\caption[]{The sphere SF corresponding to strong explosion in a
plane-layered medium with the inverse square density decrease.
a)~The half-sphere adjoined to the upper leading point $z_2 (t)$
moving towards the density decrease. b)~The surface of the
sphere segment adjoined to the lower leading point $z_1 (t)$.
c)~The surface of the spherical layer complementing the SF to the
sphere. This part of the solution contrary to a) and b)
corresponds to $\mu\ne0$ (the region of the ``interference''
of waves). The parts of the SF corresponding to the solutions in
different regions (see text) are shifted here each off another.}
\label{fig1}
\end{figure}

As mentioned above, this case is important for the astrophysical
applications (outer parts of coronas, the Solar wind region,
etc.). However, it also gives us a clear example of how to express
the solution via supplementary functions.  At $ \varphi(z) =1/z^2$
$(z_0=1)$ the integrals are expressed
via elementary functions but, as seen, separate regions in the
$z,y$ plane should be considered to provide positiveness of the
expressions under the square root. Let $z$ satisfy the
inequality $z_2  > z  > z_{\rm ex}$. In this region adjacent to the
leading point of the SF we obtain ($\mu=0$):
\[
\xi ^2\varphi >1,\quad \xi \mid _{z_{\rm ex}}=z_{\rm ex},\quad ( z_0=1) ,
\]
\begin{equation}
\label{2.2}
r=\sqrt{\xi ^2-1}-\sqrt{\xi ^2-z^2},\quad ( z>1) ,
\end{equation}
\begin{equation}
\label{2.3}
\xi ^2=\frac{z^2( 1-e^{2y}) ^2}{4e^{2y}( e^y-z)
(z-e^{-y}) }.
\end{equation}
Here
\begin{equation}
\label{2.4}
z_2=e^y,\quad z_1=e^{-y},\quad z_{\rm ex}=\cosh y,\quad r\mid
_{z_{\rm ex}}=\sinh y .
\end{equation}
Calculating
\begin{equation}
\label{2.5}
\xi ^2-1=\frac{[ z( 1+e^{2y}) -2e^y]^2}{
4e^{2y}( e^y-z) ( z-e^{-y}) } \quad ,
\end{equation}
and
\begin{equation}
\label{2.6}
\xi ^2-z^2=\frac{z^2( 1+e^{2y}-2ze^y) ^2}{
4e^{2y}( e^y-z) ( z-e^{-y}) }
\end{equation}
we find that
\begin{equation}
\label{2.7}
z>z_{\rm ex}, \quad
\sqrt{\xi ^2-z^2}=\frac{z( z-z_{\rm ex}) }{\sqrt{( z_2-z)
( z-z_1) }}>0,\quad \xi \mid _{z_{\rm ex}}=z_{\rm ex}.
\end{equation}
\begin{equation}
\label{2.8}
z>z_{\rm ex}^{-1}, \quad
\sqrt{\xi ^2-1}=\frac{zz_{\rm ex}-1}{\sqrt{( z_2-z) (
z-z_1) }},\quad \xi \mid _{z_{\rm ex}^{-1}}=1.
\end{equation}
Substituting these relations into Eq.\ (\ref{2.2}) we get an
extremely simple solution:
\begin{equation}
\label{2.9}
r=\sqrt{( z_2-z) ( z-z_1) }\quad , \quad
z_2  > z  > z_{\rm ex}.
\end{equation}
Apparently, it is the equation of the hemisphere hanging on the
upper leading point $z_2$ with the centre at $z_{\rm ex}$ and radius
$(z_2-z_1)/2$ (Fig.~\ref{fig1}a).  Note that in this region really $\xi >
z$ (\ref{2.6}). At $z_1<z<z_{\rm ex}^{-1}<1$ with $\xi^2$ from
(\ref{2.3}) we have
\begin{equation}
\label{2.10}
r=\sqrt{\xi ^2-z^2}-\sqrt{\xi ^2-1},
\end{equation}
\begin{equation}
\label{2.11}
\sqrt{\xi ^2-z^2}=\frac{z( z_{\rm ex}-z) }{\sqrt{( z_2-z)
( z-z_1) }}\quad,\quad
\sqrt{\xi ^2-1}=\frac{1-zz_{\rm ex}}{\sqrt{( z_2-z) (
z-z_1) }},
\end{equation}
and, finally, (Fig.~\ref{fig1}b) the solution reduced to the same form
\[
r=\sqrt{( z_2-z) ( z-z_1) },\quad zz_{\rm ex}<1
\]
describes the segment surface of the same sphere leaning against the lower
leading point $z_1$. The square root singularities of $\xi$ at the leading
points are the general property for an arbitrary $n$ substantially used in
\cite{kp6}.

\subsection*{Intermediate region}

\begin{figure}
\centerline{\epsfbox{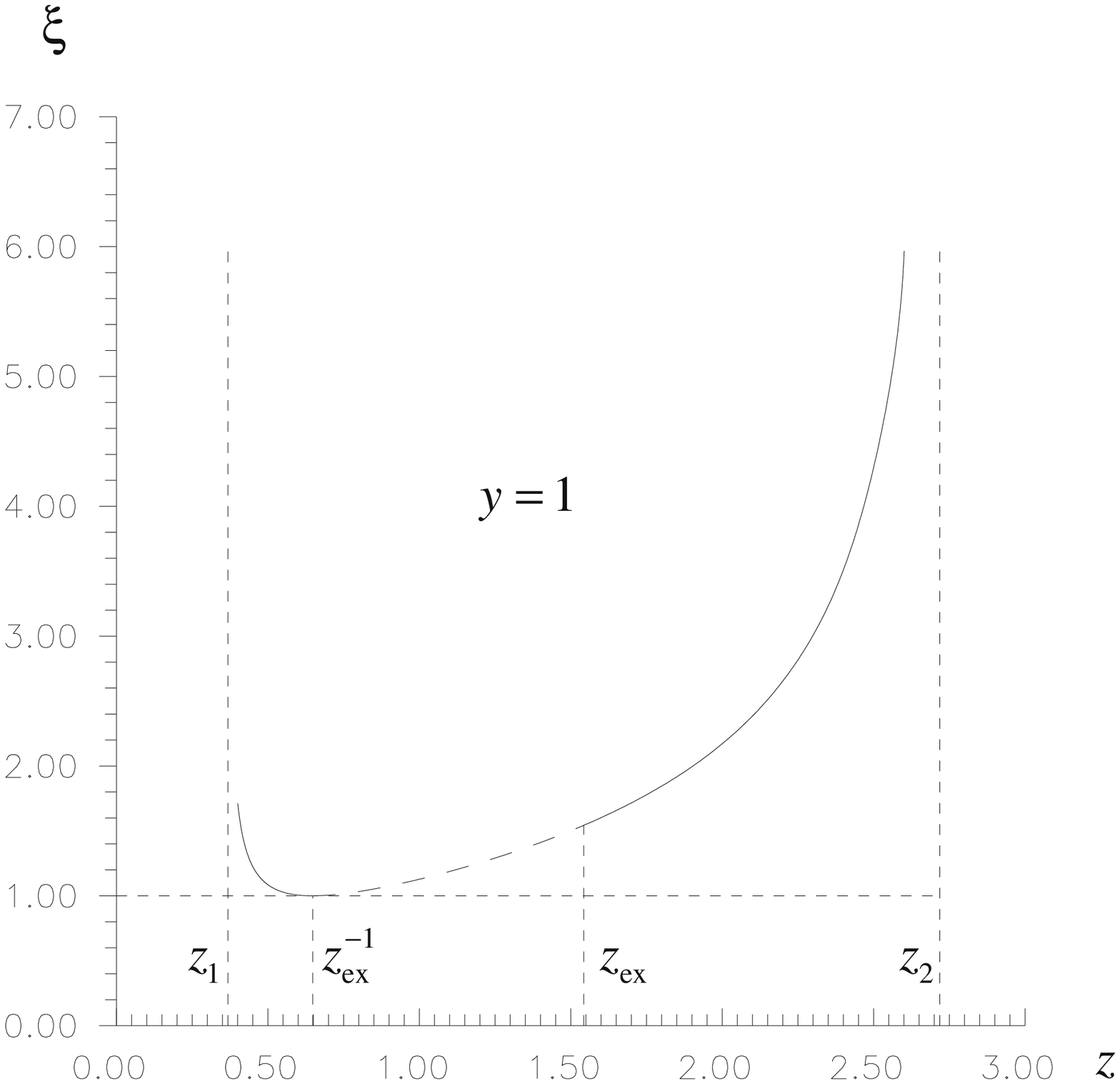}}
\vskip1cm
\centerline{\epsfbox{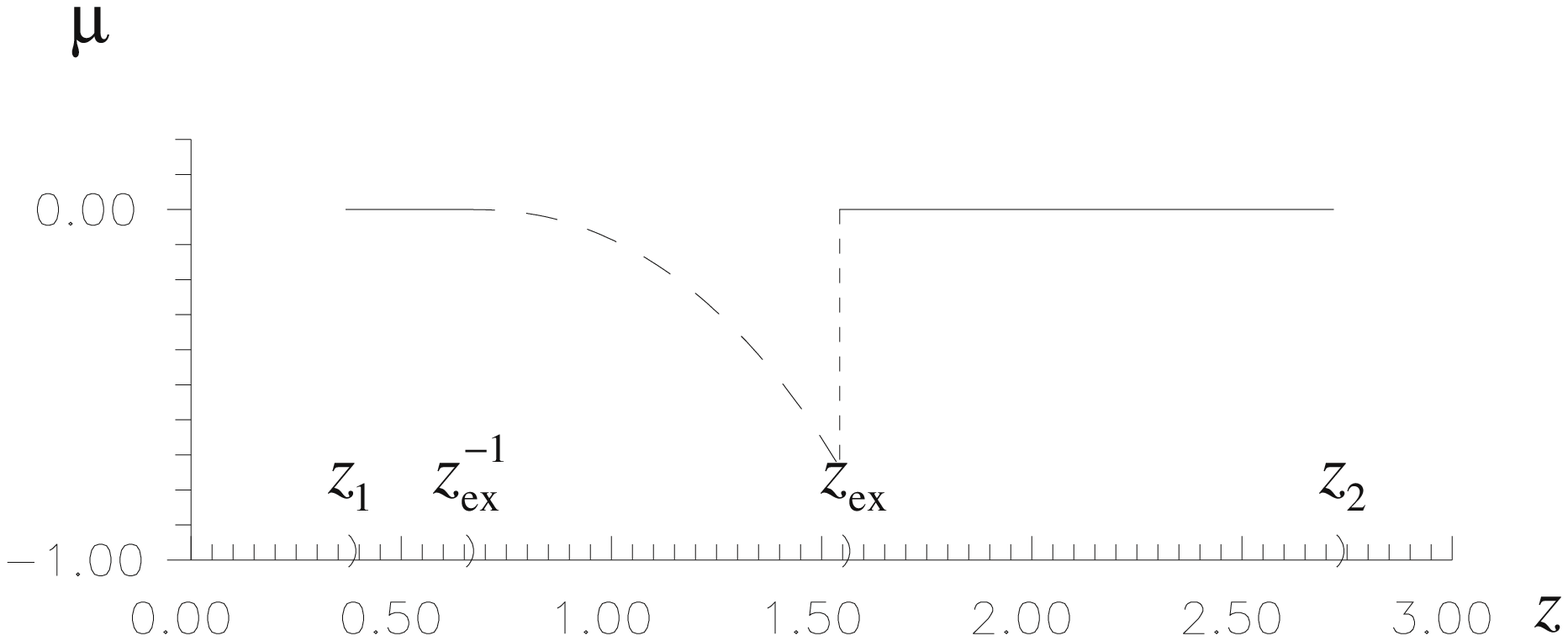}}
\caption[]{The supplementary functions $\xi$ and $\mu$
parameterising the solution for the SF. The square-root
singularities of $\xi$ at the leading points arise for an
arbitrary $n$.}
\label{fig2}
\end{figure}

Here we return to the initial general form of the solution with the
arbitrary function which is found from matching conditions at the boundary
between the regions (compare to Silich \& Fomin \cite{sf})
\begin{equation}
\label{3.1}
r\mid _{z_{\rm ex}-0}=r\mid _{z_{\rm ex}+0},\quad ( z_{\rm ex}=\xi \mid
_{z_{\rm ex}})
\end{equation}
\begin{equation}
\label{3.2}
\mu ( \xi ) =2\sqrt{\xi ^2-1}-2\xi \ln \left\{ \xi +
\sqrt{\xi ^2-1}\right\}
\end{equation}
\begin{equation}
\label{3.3}
\mu ^{\prime }( \xi ) =-2\ln \left\{ \xi +\sqrt{\xi ^2-1}\right\}
\end{equation}
Substituting these relations into
\begin{equation}
\label{3.4}
0=\frac{\partial r}{\partial \xi }=\int_1^z\frac{\xi \varphi \,dx}{\sqrt{\xi
^2\varphi -1}}+y+\mu ^{\prime }( \xi )
\end{equation}
we get the equation for $\xi$ at $ z_{\rm ex}>z>z_{\rm ex}^{-1} $
\begin{equation}
\label{3.5}
y=\ln \left| \frac{\xi +\sqrt{\xi ^2-z^2}}z( \xi +\sqrt{\xi
^2-1}) \right| .
\end{equation}
Using the definition
\begin{equation}
\label{3.6}
\xi \equiv \cosh \alpha
\end{equation}
we obtain
\begin{equation}
\label{3.7}
\alpha =\ln \sqrt{\frac{z_2( zz_2-1) }{( z_2-z) }},
\end{equation}
\[
\alpha \mid _{z_{\rm ex}}=y,\quad \alpha \mid _{z_{\rm ex}^{-1}}=0.
\]
Thus, the value $\xi$ (Fig.~\ref{fig2}a)
in this region is given by the relation
\begin{equation}
\label{3.8}
\xi \mid _{z_{\rm ex}}=\cosh y=z_{\rm ex},\quad \xi \mid _{z_{\rm ex}^{-1}}=1.
\end{equation}
\begin{equation}
\label{3.9}
\xi =\cosh \ln \sqrt{\frac{e^y( e^yz-1) }{e^y-z}}
,\quad z_{\rm ex}>z>z_{\rm ex}^{-1}.
\end{equation}
It is seen that
\begin{equation}
\label{3.10}
\mu _{z_{\rm ex}^{-1}}=\mu _{z_{\rm ex}^{-1}}^{\prime }=0,
\end{equation}
\[
\mu \neq 0,\quad \mu ^{\prime }\neq 0\quad (z=z_{\rm ex}).
\]
Therefore $\mu$ is continuous at the lower boundary but together with
the first derivative it jumps at the upper one (Fig.~\ref{fig2}b).
\begin{equation}
\label{3.11}
\mu ( \xi _{\rm ex}) =2( \sinh y-y\cosh y) ,\quad \mu
^{\prime }( \xi _{\rm ex}) =-2y;\quad \xi _{\rm ex}=\xi (
z_{\rm ex}) .
\end{equation}

Hence, in the region $z_{\rm ex} > z > z_{\rm ex}^{-1}$
the expression can be written for $\xi$
\begin{equation}
\label{3.12}
\xi =\frac 12\left\{ \sqrt{\frac{e^y( e^yz-1) }{e^y-z}}
+\sqrt{\frac{e^y-z}{e^y( e^yz-1) }}\right\},
\end{equation}
or
\begin{equation}
\label{3.13}
\xi =\frac{z\sinh y}{\sqrt{( z_2-z) ( z-z_1) }}.
\end{equation}
The following relations also can be found
\begin{equation}
\label{3.14}
\xi = \frac{z( z_2-z_1) }{2\sqrt{( z_2-z) (
z-z_1) }},
\end{equation}
\begin{equation}
\label{3.15}
\xi ^2-1=\frac{( z\cosh y-1) ^2}{( z_2-z)
( z-z_1) },\quad \xi ^2-z^2=\frac{z^2( \cosh y-z) ^2}{
( z_2-z) ( z-z_1) }.
\end{equation}

\begin{figure}
\centerline{\epsfbox{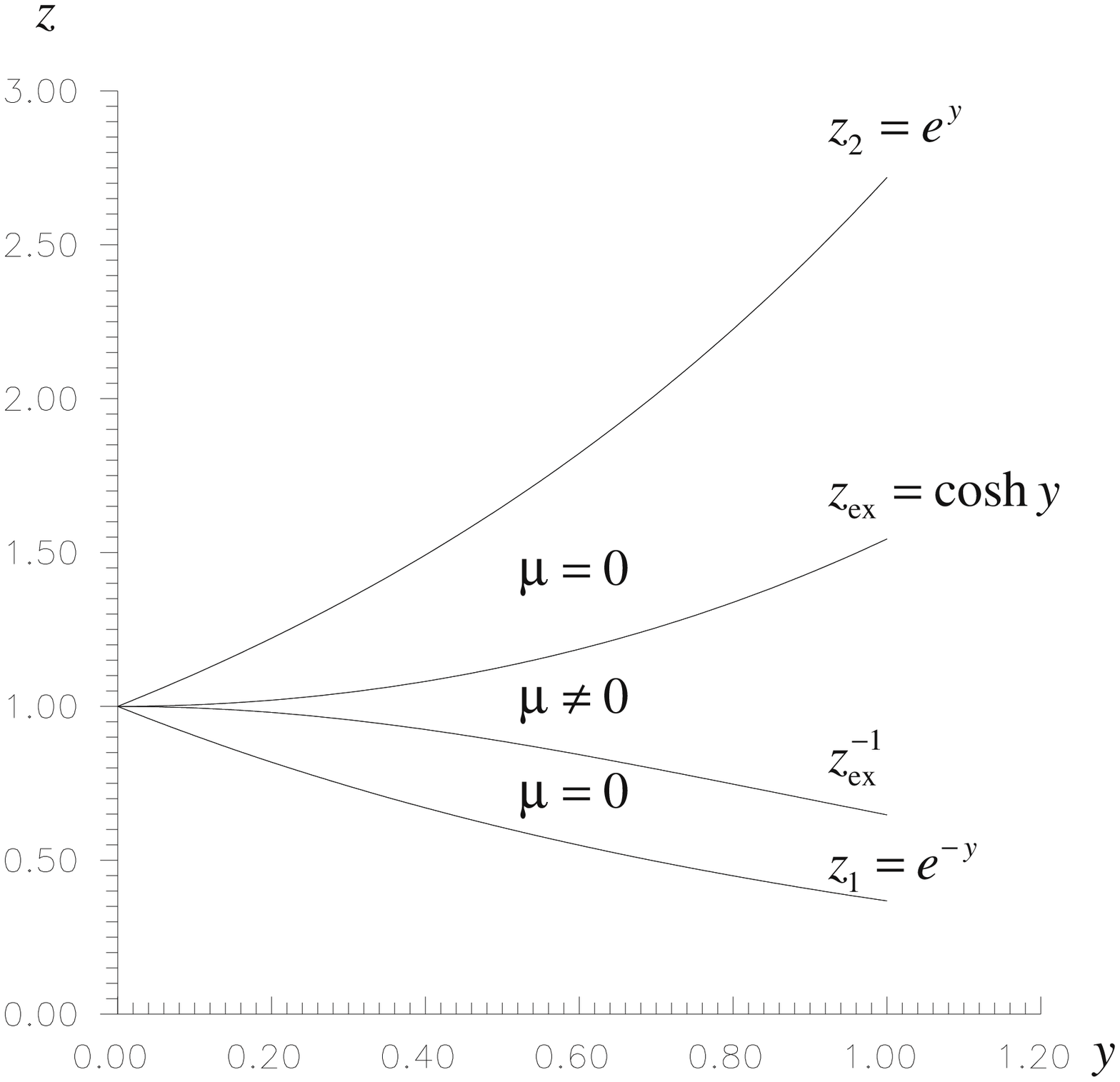}}
\caption[]{The dependence of specific points of the SF on ``time''
$y$. It is seen that the region $\mu\ne0$ vanishes at $y\to0$. At
every moment the solution is a sphere with the centre at $z=\cosh
y$ and radius growing as $\sinh y$ ($z_0=1$ is accepted).}
\label{fig3}
\end{figure}

Finally, in the intermediate region (Fig.~\ref{fig1}c) the following
formula is valid
\begin{equation}
\label{3.16}
r=\sqrt{\xi ^2-z^2}+\sqrt{\xi ^2-1}
\end{equation}
and according to (\ref{3.15}),
\begin{equation}
\label{3.17}
r=\sqrt{( z_2-z) ( z-z_1) }.
\end{equation}
Thus, we get the complementary surface of a spherical layer, which
together with another segments above forms a complete sphere. It
is a complete and extremely simple solution of the problem. We
also present it in an explicit form indicating temporal
dependencies of the envelope radius and centre:
\begin{equation}
\label{10}
r=z_{0}[(\exp \{y/z_{0}\}- z/z_{0})(z/z_{0} - \exp \{-y/z_{0}\})]^{1/2}.
\end{equation}
The solution can be also presented in the canonical form:
\begin{equation}
\label{11}
r^{2}+(z-z_{0}\cosh(y/z_{0}))^{2} = z^{2}_{0}\sinh^{2}(y/z_{0}).
\end{equation}
At every moment the solution is a sphere with the centre at
$z=z_{0}\cosh(y/z_{0})$
and the radius growing as
$z_{0}\sinh(y/z_{0})$.
The dependence of the real time $t$ on the Kompaneets variable $y$
\[
t \propto \int^y \, dy \, \sinh^{3/2}y
\]
can not be exactly expressed via elementary functions even for
this simplest case but the asymptotical relation is obviously
simpler. Such a perfect form of the solution in an inhomogeneous
medium is very surprising. It is much closer to the Sedov solution
than to the Kompaneets one in full accordance with the
comparatively slow density dependence on the coordinate (see
footnote 1).  Nevertheless, the velocity decrease of the leading
point at large time ($z_{2}\gg z_{0} )$) $dz_{2}/dt \propto
t^{-1/3}$ is slower compared to the homogeneous medium $dz_{2}/dt
\propto t^{-3/5}$.  Note that at large $y\gg 1$
\[
z_{\rm ex}\approx \frac{z_2}2\, ,\quad z_{\rm ex}^{-1}\approx 2z_1.
\]
At small
$y\ll 1$
\[
z_{\rm ex}\approx 1+\frac{y^2}2\, , \qquad z_{\rm ex}^{-1}\approx 1-\frac{
y^2}2.
\]
When approaching to the point of explosion, the intermediate
region gets narrow and disappears at all (Fig.~\ref{fig3}).

\subsection*{Comparison to results of the conform transformation
for radially stratified medium}

Here we compare the results discussed above to the brilliant
solution by Korycansky \cite{kor} for off-centre explosion in a
radially stratified medium. In the Kompaneets approximation the
equation $\chi ( R,y) =\theta$ for the SF can be
written as:
\begin{equation}
\label{4.1}
{\left (\frac {\partial \chi }{\partial y}\right )}^2=
\frac {1}{\psi (R)}
\left[ {\left (\frac {\partial \chi }{\partial R}\right )}^2+\frac
1{R^2}\right],\quad \psi (R) \equiv \frac{\rho(R)}{\rho _0 }
\end{equation}
Here $R,\theta $ are the spherical coordinates, $R=a,\theta =0,(y=0) $
are the coordinates of the explosion. At $ \rho ( R)
=\rho _0( \frac aR) ^n$, the substitution
\cite{kor} $ \zeta =(\frac {R}{a})^\alpha$, $\phi =\alpha
\chi$, $\alpha =\frac{2-n}2$ leads to the equation in an
effective homogeneous medium
\begin{equation}
\label{4.3}
{\left
(\frac {\partial \phi}{\partial y}\right )}^2= \frac
1{y_c^2}\left[  {\left (\frac {\partial \phi }{\partial \zeta
}\right )}^2+\frac 1{\zeta ^2}\right] ,\quad y_c=\frac{2a}{|
2-n| }.
\end{equation}
The solution of \cite{kor} is given by the relations:
\begin{equation}
\label{4.4}2\zeta \cos \phi =1+\zeta ^2-x^2,\quad x \equiv \frac {y}{y_c},
\end{equation}
\begin{equation}
\label{4.5}
2\left (\frac{R}{a}\right )^{\frac{2-n}2}\cos \frac{( 2-n)
\theta }2=1+\left (\frac{R}{a}\right )^{2-n}-
\left[ \frac{y( 2-n)}{2a}\right ]^2.
\end{equation}
For $n=2$ the following special solution exists \cite{kor}:
\begin{equation}
\label{4.6}
\theta ^2+\ln^2\frac Ra=\left (\frac {y}{a}\right )^2.
\end{equation}
It can be easily found using the transformation from the spherical case
\begin{equation}
\label{4.7}
{\left (\frac {\partial \chi }{\partial y}\right )}^2=\frac 1{a^2}\left[
R^2 {\left (\frac {\partial \chi }{\partial R}\right )}^2+1\right]
\end{equation}
to the effective plane one with the help of the substitution:
$\eta =\ln R$, $y\to ya$:
\begin{equation}
\label{4.8}
{\left (\frac {\partial \chi }{\partial y}\right )}^2= {\left (\frac
{\partial
\chi }{\partial \eta }\right )}^2+1\Longrightarrow \chi ^2+\eta ^2=y^2.
\end{equation}
The same result can be obtained from the general formula (\ref{4.5})
in the limit:
\begin{equation}
\label{4.9}
\eqalign{
{R}{a}^{(2-n)/2}\to \exp \frac{2-n}2\ln \frac Ra
\to\\ 1+\frac{2-n}2\ln \frac Ra+\frac {1}{2}
\left(\frac {2-n}{2}\right)^2 \ln ^2\frac Ra+\dots\quad ( 2-n\to
0)}
\end{equation}

It can be easily verified that main parameters of the surface
(\ref{4.6}) are very close to that of the sphere discussed above.
Correspondence to the case of homogeneous medium can be verified
using the transformation to the cylindrical coordinates
\begin{equation}
\label{4.10}
\left(\frac {\tilde \rho }{z}\right)^2=\tan^2\sqrt{(\frac
{y}{z_0})^2-\ln^2\frac{\sqrt{{\tilde \rho }^2+z^2}}{z_0}},\quad
{\tilde \rho }=R\sin \theta, \quad z=R\cos \theta.
\end{equation}
At small time
\begin{equation}
\label{4.11}
\left(\frac {\tilde \rho }{z}\right)^2=\frac{( y/z_0 )^2-\ln
^2z/z_0}{1+\ln z/z_0},\quad {\tilde \rho } \ll z,z_0,
\end{equation}
it becomes the Sedov solution in ordinary variables:
\begin{equation}
\label{4.12}
{\tilde \rho }^2+( z-z_0) ^2=y^2.
\end{equation}

\subsection*{Correspondence of solutions for plane-layered and
radially stratified media and a new solution for SF in the case of
density singularity on the finite radius}

The Korycansky transformations can be generalised. The following
substitution of variables
\begin{equation}
\label{a1}
\frac z{z_0}=\ln \frac Ra  \qquad (R=a\exp \frac z{z_0})
\end{equation}
transforms the Kompaneets equation for $r(z,y)$ in a plane-layered medium
(\ref{3}) (for cylindrical coordinates $z,r $) to that for $\chi (R,y)$ in a
radially stratified medium (\ref{4.1}) (for spherical coordinates $R,\chi $).
This substitution also transforms the density dependence $\varphi (z)$ in a
plane-layered medium to $\psi(R)$ in spherical one according to the relation
\begin{equation}
\label{a2}
\psi (R)=\varphi (z_0\ln \frac Ra)\frac{z_0^2}{R^2}
\end{equation}
and vice versa:
\begin{equation}
\label{a3}
\varphi (z)=\psi (R)\frac{R^2}{z_0^2}=({a}/{z_0})^2\psi
(a\exp \frac z{z_0})\exp \frac{2z}{z_0}.
\end{equation}

Corresponding solutions of the Kompaneets equations are related as
\begin{equation}
\label{a4}
r=z_0\chi.
\end{equation}

Note that in this case the exponential density dependencies
\begin{equation}
\label{a5}
\varphi (z)=\exp \{-\beta z\}
\end{equation}
are transformed to the power ones
\begin{equation}
\label{a6}
\psi (R)=\frac{z_0^2}{R^2}\exp \{-\beta z_0\ln \frac Ra\}\propto
R^{-(\beta z_0+2)}.
\end{equation}
The equivalent height of the atmosphere for the plane case
determines the exponent for the power law in the radially-stratified
medium. Particularly, this fact clarifies existence of
the solution sequence in the Korycansky case \cite{kor}, which therefore
are connected with Kompaneets solutions \cite{kom}.

In the special case
\[
\varphi (z) \propto 1/z^2
\]
the exact solution for which is constructed above (see (\ref{10})) the law
\begin{equation}
\label{a7}
\psi (R) \propto 1/R^2\ln^2(R/a)
\end{equation}
gives the density dependence very close to the inverse-square law
at large distances and having a singularity at the surface of the
sphere with the radius $a$.

Correspondingly, there is a new formal solution with a fixed point of explosion
$R=ae$:
\begin{equation}
\label{a8}
\chi (R)=\sqrt{(\ln (R/a)-z_1/z_0)(z_2/z_0-\ln (R/a))},
\end{equation}
where $z_{1,2}(y)/z_0$ are defined by the formula (\ref{2.4}).

The substitution (\ref{a1}) is also suitable for computations
relating to the off-centre explosions in radially stratified media.

\subsection*{Conclusion}

The results obtained are in good agreement with the known
self-similar solutions \cite{p,h,sak}. These results can be useful
not only for the Solar corona but for another astrophysical
situations (compare to \cite{ch,gv}) where the inhomogeneity of
media is important and, moreover, the inverse-square law of the
density decrease occurs very often\footnote{For example, in the
solar corona it is confirmed by measurements \cite{ma} in the
range from 4 to 200 solar radii. (The authors are grateful to N.~Lotova
for this information.) The similar dependence is
observed for outer atmospheres of young stars and for the
periphery of molecular clouds (see references in the review
\cite{l}).}, for example, for Supernovae in the periphery of
gigantic clouds, for outbursts in nebulae \cite{l}, for varieties
of galactic fountains \cite{ks}, for galaxies with star formation
outbreaks \cite{bs} and also for more exotic cases such as the
Supernova in the NGC 6946 nearby another Supernova (i.e., in its
wind region) \cite{usp} recently discovered by the Hubble Space
Telescope or for galaxy collisions.

Particularly, the off-centre location of pulsars compared to
Supernova Remnants can be naturally explained by displacement of
the envelope towards the direction of the density decrease without
attracting the idea of the non-symmetrical explosion and recoil.
The solution in the medium with the singularity of the density at
a sphere surface of a finite radius describes the bypass of the
photosphere by the SF and is of great interest for the Sun and
stars. The asymptotical solutions for the leading point reveal
possible non-monotonic behaviour of the SF velocity at
the monotonic density decrease (see \cite{kp7}). The solutions
obtained admits an interesting generalisation if the constant
energy $E_0$ is replaced by the function of time $E(t)$ which
allows to take into account losses \cite{ks} as well as an energy
supply (see notice in \cite{kp7}).

\subsection*{Acknowledgment}

V.~M.~Kontorovich thanks the International Soros Science
Educational Program of the International Renaissance Foundation
grant SPU042029 for support, and S.~F.~Pimenov thanks Russian
Astronomical Council for supporting his work by a grant.

\subsection*{Appendix~1}

The Kompaneets equation for the SF is derived by equating the
normal component of the shock front velocity $D=(p/(\rho - \rho
^{2}/\rho ^\prime ))^{1/2}$ obtained from boundary conditions to
its formal one $-(\partial f/\partial t)/|\nabla f|$.
Here $p=(\gamma -1)\lambda E_{0}/V$ is the pressure behind the
shock front, $\gamma $ is the adiabatic exponent, $E_{0}$ is the
energy of the explosion, $\rho $ and $\rho ^\prime $ are densities
ahead and behind the shock front, $\rho ^\prime /\rho =(\gamma
+1)/(\gamma -1)$, $V(t)$ is the volume of the envelope $V(t)=\pi
\int r^{2}dz$, with the integral taken over the part of $z$
axis, restricted by the envelope.  The constant $\lambda $, is
of about to 2--3 and accounts the deviation of the pressure
near the shock front from the average value, $f(r,z,t)=0$ is the
shock front equation.

\subsection*{Appendix~2}

Some notes should be made about the leading point behaviour for $\varphi
(z)=(z/z_{0})^{-n}$ at $n$ differing from 2. Relations (\ref{0.1}),
(\ref{0.3}) and conditions $r(z_{1,2})=0 $ of intersection of axis $z$ by
the SF give the following equations for the values $z_{1,2}(y)$:
\begin{equation}
\label{12*}
y = \int _{z_0}^{z_2} dz^\prime \Bigl(\varphi (z^\prime )\Bigr)^{1/2} =
\int _{z_1}^{z_0} dz^\prime \Bigl(\varphi (z^\prime )\Bigr)^{1/2}.
\end{equation}
In the vicinity of the point $z_2$ the value $\xi $ is large
and an expansion by $1/\xi \ll 1$ results in the following
expression for $\xi ^{2}$ \cite{kp6}:
\begin{equation}
\label{x}
\xi ^2\approx \left[ 2\left( z_2-z\right) \left( \varphi
(z_2)\right) ^{1/2}\right] ^{-1}{\int^{z_2}_{z_0}}du\left(
\varphi (u)\right) ^{-1/2}
\end{equation}
Correspondingly,
\begin{equation}
\label{a}
r\approx (1/\xi (z,y))\cdot{\int^z_{z_0}}du\left(
\varphi (u)\right) ^{-1/2},\qquad z\to z_2(y).
\end{equation}
Thus, the function $\xi$ has a square-root
singularity with accordance to (\ref{2.3}). These relations can be used to
estimate the volume $V(t)$ required to describe motion of the SF in terms
of real time $t$. For $z_2$ we obtain:
\begin{equation}
\label{18*}
z_2(t) = \frac{z_0}{\left(1 - \sign(n-5) t/t_{*}\right)^{2/(n-5)}},
\end{equation}
\begin{equation}
\label{19*}
t_{*} = \sqrt {16 \pi z_0^5 \rho _0
/ (n+2)(n-5)^2E_0\lambda
(\gamma ^2 - 1)}.
\end{equation}

As it follows from the relation (\ref{12*}) in the case of $n>5$
\cite{kor,kp6} ``disastrous'' acceleration takes place for the
mentioned part of the SF which approaches the infinity $z_{2}=
\infty $ for a finite time.  The expression for the velocity can
be written as
\begin{equation}
\label{16*}
dz_{2}/dt \propto
(t_{*}- t)^{(n-3)/(5-n)}.
\end{equation}
The specific time
$t_{*}$ of ``disastrous'' acceleration is
\begin{equation}
\label{17*}
t_{*} \approx ( z^{5}_{0} \rho _{0}/E_{0}
)^{1/2}
\end{equation}
(see also \cite{kp6}). Despite the fact
that we used here the solution valid only in the vicinity of the
leading point the relations (\ref{18*}) and (\ref{16*}) are enough
adequate to the phenomena. However, the problem still remains of
finding the volume and some features are lost when using such
calculations.

\end{document}